\documentclass[prd,a4paper,floatfix,
twocolumn,
]{revtex4-1}

\usepackage{amsmath}
\usepackage{amsfonts}
\usepackage{amssymb}
\usepackage{graphicx}
\usepackage[raggedright,tight]{subfigure}
\usepackage[utf8]{inputenc}
\usepackage[american]{babel} 
\usepackage{slashed} 
\usepackage{color}
\usepackage{hyperref}
\usepackage{bm}
\usepackage[percent]{overpic}

\definecolor{oucrimsonred}{rgb}{0.6, 0.0, 0.0}
\definecolor{persianblue}{rgb}{0.11, 0.22, 0.73}
\definecolor{forestgreen}{rgb}{0.13,0.35,0.13}
 \hypersetup{colorlinks, citecolor=oucrimsonred, linkcolor=persianblue, urlcolor=oucrimsonred}
\newcommand{\be}{\begin{equation}}
\newcommand{\ee}{\end{equation}}
\newcommand{\bea}{\begin{eqnarray}}
\newcommand{\eea}{\end{eqnarray}}

\newcommand{\eq}[1]{equation~\eqref{#1}}
\newcommand{\secref}[1]{section~\ref{#1}}
\newcommand{\ssecref}[1]{subsection~\ref{#1}}
\newcommand{\dph}{\ensuremath{\bar{\gamma}}}
\newcommand{\dps}{\ensuremath{a}}

\begin{document}


\title{$Z$-boson decays into an invisible dark photon at the LHC, HL-LHC  and future lepton colliders}

%
%
\author{M.\ Cobal$^{\star \S}$}
\author{C. De Dominicis$^{\ast}$}
\author{M.\ Fabbrichesi$^{\dag}$}
\author{E.\ Gabrielli$^{\circ \dag \diamond }$}
\author{J. Magro$^{\circ \S}$}
\author{B. Mele$^\ddag$}
\author{G. Panizzo$^{\star \S}$}

 \affiliation{$^\star$Dipartimento Politecnico di Ingegneria ed Architettura, University of Udine, Via della Scienze 206, 33100 Udine, Italy }
 \affiliation{$^\S$INFN, Gruppo Collegato di Udine}
 \affiliation{$^{\ast}$Subatech, CNRS/IN2P3, IMT-Atlantique, Universit\'{e} de Nantes, France}
 \affiliation{$^{\dag}$INFN, Sezione di Trieste, Via  Valerio 2, 34127 Trieste, Italy }
 \affiliation{$^\circ$Physics Department, University of Trieste, Via  Valerio 2, 34127 Trieste, Italy }
\affiliation{$^{\diamond}$NICPB, R\"avala 10, Tallinn 10143, Estonia }
\affiliation{$^\ddag$INFN, Sezione di Roma, P.le Aldo Moro 2, 00185 Roma, Italy }

\date{\today}

\begin{abstract}
We study the decay of the $Z$ vector boson into a photon and a massless (invisible) dark photon in high-energy collisions.  The  photon  can be used as  trigger for the event, while the dark photon  is detected  indirectly as missing  momentum in the event final state.  We investigate the possibility of  searching for such a dark photon  at the LHC, HL-LHC and   future lepton colliders, and compare the respective sensitivities. As expected, the best result is found for the lepton colliders  running at the $Z$ mass, FCC-ee and CEPC, with a final sensitivity to  branching ratios of order $O(10^{-11})$. We also discuss how to use the photon angular distribution of the events in lepton collisions to discriminate between the dark photon and a pseudo-scalar state like the axion.
\end{abstract}

\maketitle


\section{\label{sec:intro} Motivations and synopsis}

The two-body $Z$ boson decay into an isolated  photon~$\gamma$ and 
a stable (or meta-stable) neutral particle beyond the Standard  Model  (SM),   effectively  coupled  to  the  $Z$ and  $\gamma$ gauge bosons, can give rise to a peculiar experimental signature at high-energy colliders.


The kinematical properties of the detected photon and of the neutral particle, which are both monochromatic in the rest frame of the decaying $Z$, makes this process quite attractive in the search for new physics effects---most notably in the case of lepton colliders, for which the monochromatic photon energy is smeared only by \textit{Bremsstrahlung} radiation and detector effects. This feature is  lessened at hadron colliders because of the additional challenges in reconstructing the  rest frame of the $Z$-boson, due to the characteristic uncertainties in  the transverse missing momentum measurement. 
 
 What are the possible candidates for such a neutral particle?
 

 Within supersymmetric  theories, neutral states can be either fermions (neutralinos and gluinos), which, like the SM neutrinos, would only contribute to  $Z$ three-body decays, or scalars (sneutrinos), which are too heavy  to be produced in the $Z$-boson decay.  

Other theoretical frameworks remain open.
Here we consider the case of the \textit{dark photon}, which 
 is a particle belonging to a \textit{dark} sector---a sector comprising particles interacting only feebly  with SM states, and fashioned as a generalization of dark matter (see e.g.  \cite{Fabbrichesi:2020wbt} for a recent review).

A  dark photon $\dph$ is the gauge boson of a  $U(1)_D$ group under which dark matter as well as all other dark particles are charged.
In particular, a \textit{massless} dark photon (corresponding to an unbroken
$U(1)_D$ gauge group)  does not couple directly to  the SM currents, but only through a dipole-like operator of dimension-five~\cite{Hoffmann:1987et,Dobrescu:2004wz}, a distinguishing characteristic that makes the massless case very different from the  massive one. As far as its phenomenology is concerned, the massive dark photon interacts (via mixing effects)  like  a SM photon. 
Since the decay of the $Z$ boson into two photons  is  forbidden by the Landau-Yang theorem~\cite{Landau:1948kw,PhysRev.77.242}, a $Z$ boson cannot decay into a photon and a massive dark photon via mixing. 
On the other hand, 
 the decay of the $Z$ boson into a photon and a massless dark photon is mediated by  a one-loop diagram of SM fermions containing the $\dph$ dark dipole operator  and the usual electromagnetic vertex for the emission of the SM photon.  Then the  Landau-Yang theorem does not apply to this case because the interaction vertices of the dark and the ordinary photon are distinguishable. A  branching ratio (BR) for $Z\to \gamma \dph$ between  $O(10^{-9})$ and $O(10^{-11})$ can be expected~\cite{Fabbrichesi:2017zsc}. 

In the present $Z$ decay channel,  the missing energy  could be carried away by  neutral bosons other than the dark photon. For example, an axion-like 
particle (ALP) has been considered and found to have a BR as large as $O(10^{-4})$~\cite{Bauer:2017ris,Kim:1989xj}.
Also more exotic cases have been suggested: a Kaluza-Klein graviton in models of large extra-dimensions (with a BR around $O(10^{-11})$)~\cite{Allanach:2007ea}  and an unparticle (for which a BR as large as $O(10^{-6})$ is possible)~\cite{Cheung:2008ii}. The signature in the latter two cases will be different since the photon is not resonant due to the continuum spectrum of the missing mass. The same signature is shared by the irreducible background $Z\to \gamma \nu \bar{\nu}$~\cite{Hernandez:1999xn}.



At the experimental level, the process  
\be e^{+} e^{-} \to Z \to \gamma + X^0 \, ,
\ee
where $X^0$ stands for  undetected neutral particles, was explored at 
the Large Electron-Positron Collider (LEP)  and a limit of $10^{-6}$ at the 95\% CL was found \cite{Abreu:1996vd} for the corresponding BR when considering a massless $X^0$ in the final state.

In this paper,  we investigate the possibility of  searching for such a massless dark photon  in $Z$ decays at the Large Hadron collider (LHC), the High Luminosity LHC (HL-LHC), and   future lepton colliders, and compare the respective sensitivities. 

The paper is organized as follows.
 In \secref{sec:TH}, we  provide the theoretical framework for the effective $Z\gamma \dph$ couplings, and the expression for the corresponding effective
Lagrangian.
In  \secref{sec:lhc}, we study, through Monte Carlo (MC) simulations, the upper limit on the BR of the $Z$ decay into a photon and a dark photon, that can be reached using data collected at the LHC. Performing a simple extrapolation from the result obtained for the LHC, we also estimate  the  limit on the BR which can be obtained at the HL-LHC. 
In \secref{sec:Future Machines}, we extend the study to Future Circular Colliders, namely the FCC-ee  and the CEPC. At the electron-positron colliders, as expected, the lower level of background compared to hadron colliders and the production of large samples of $Z$ bosons provide the most stringent limit on the BR. In \secref{sec:spin}, assuming that the decay has been observed, we look at the angular distribution of the events to establish the spin  of the particle carrying away  the missing energy (see~\cite{Comelato:2020cyk} for a discussion of the spin dependence of the signature), and determine a lower bound on the number of events needed to distinguish between the case of the spin-1 dark photon from a spin-0 pseudo-scalar.
In \secref{sec:concl}, we give our conclusions.

\section{\label{sec:TH}Theoretical framework}
We consider  the effective coupling of the photon $\gamma$ and the $Z$ gauge boson to a massless dark photon \dph. The  SM fermions couple to a massless dark photon $\dph$  only through radiative corrections 
induced by loops of  dark-sector  particles. 
The starting point is thus given by  considering the leading contribution  provided by the magnetic- and electric-dipole operators, described by the effective Lagrangian 
\be
{\cal L}=\sum_f \frac{e_D}{2\Lambda} \bar{\psi}_f \sigma_{\mu\nu} \left( d^f_M + i \gamma_5 d^f_E \right)\psi_f 
B^{\mu\nu}\, ,
\label{dipole}
\ee
where the sum runs over all SM fields $\psi_f$, $B^{\mu\nu}$ and $e_D$ are the $U(1)_D$ dark photon field strength and elementary charge (we assume universal couplings), and $\Lambda$ the  effective scale of the dark sector.

The magnetic- ($d^f_{M}$)  and electric-dipole ($d^f_{E}$) factors can arise for instance at one-loop order, in the framework of a UV completion of the dark sector, in which there are messenger fields providing an interaction between SM and dark fields, as discussed, for instance,  in  \cite{Gabrielli:2016vbb}. In this case, the scale $\Lambda$ will be associated to the characteristic mass scale of the new physics running in the loop. The scale $\Lambda$ and the couplings $d^f_{M,E}$ can be considered as free parameters.

As shown in \cite{Fabbrichesi:2017zsc}, a non-vanishing effective coupling $Z\gamma\dph$ can be generated at one loop, inducing the manifestly gauge invariant effective Lagrangian 
\bea
{\cal L}^{(M)}_{eff}= \frac{e}{\Lambda}\sum_{i=1}^3 C_i {\cal O}_i(x) \, ,
\label{LeffMD}
\eea
where $e$ is the unit of electric charge, $\Lambda$ is the scale of the new physics, the dimension-six operators ${\cal O}_i$ are given by
\bea
{\cal O}_1 (x )&=&Z_{\mu\nu}\tilde{B}^{\mu\alpha} A^{\nu}_{~\alpha} \, ,\\
{\cal O}_2 (x) &=&Z_{\mu\nu}B^{\mu\alpha} \tilde{A}^{\nu}_{~\alpha} \, ,\\  
{\cal O}_3  (x) &=& \tilde{Z}_{\mu\nu}B^{\mu\alpha} A^{\nu}_{~\alpha}  \, ,
\eea
the field strengths $F_{\mu\nu}\equiv\partial_{\mu}F_{\nu} -\partial_{\nu}F_{\mu}$, for $F_{\mu\nu}=(Z,B,A)_{\mu\nu}$, correspond to the $Z$-boson ($Z_{\mu}$), dark-photon ($B_{\mu}$) and photon ($A_{\mu}$)  fields, respectively, and 
$\tilde{F}^{\mu\nu}\equiv \varepsilon^{\mu\nu\alpha\beta}F_{\alpha\beta}$ is the dual field strength.
The coefficients $C_i$ are dimensionless couplings that can be computed from the UV completion of the theory. For the case of the Lagrangian in \eq{dipole} they  are a function of the couplings $d_M^f$, the $U(1)_D$ unit of charge $e_D$
, the SM fermion masses, and the $Z$ mass~\cite{Fabbrichesi:2017zsc}.

Analogously, the  Lagrangian induced by the electric-dipole moment is 
\bea
{\cal L}^{(E)}_{eff}= \frac{e}{\Lambda}C_E {\cal O}(x)  \, ,
\label{eq:LeffED}
\eea
where the dimension-six operator is 
\bea
{\cal O} (x)=Z_{\mu\nu} A^{\mu\alpha} B^{\nu}_{~\alpha} \, ,
\eea
and the expression for the coefficient  $C_E$ for the Lagrangian in \eq{dipole} can be found   in \cite{Fabbrichesi:2017zsc}. The operators in \eq{LeffMD} and \eq{eq:LeffED} are  CP even and odd, respectively.

The  amplitudes in momentum space  for the decay $Z\to \gamma \dph$ can be found  by taking into account the effective Lagrangians in \eq{LeffMD} and \eq{eq:LeffED}; the total width is given by~\cite{Fabbrichesi:2017zsc}
\bea
\Gamma (Z\rightarrow \gamma \dph) &=&  \frac{\alpha M^5_Z }{6 \Lambda^4}
\left(|C_{M}|^2 +  |C_{E}|^2\right) , \label{BRTOT}
\eea
where  $C_{M}=\sum_i C_i$.

In the following we study $\text{BR}(Z\rightarrow\gamma\dph)=\Gamma (Z\rightarrow \gamma \dph)/\Gamma^Z_{\text{tot}}$ as the observable providing  a direct probe to the $Z\rightarrow\gamma\dph$ process, investigating its  discovery potential both at present and future hadron and lepton colliders.

\section{\label{sec:lhc}Hadron colliders}

The LHC \cite{Evans_2008} is a circular superconducting proton-synchrotron situated at the CERN laboratory, which accelerates and collides protons at a center-of-mass (CM) energy of 13 TeV. LHC hosts two general purpose experiments which study the collision products: ATLAS \cite{Aad:2008zzm} and CMS \cite{Chatrchyan:2008aa}. In this paper we assume an ATLAS-like detector, simulating events at 13 TeV with a total integrated luminosity of $140~\text{pb}^{-1}$, a choice that reproduces conditions similar to those obtained during Run-2 at the LHC.
The HL-LHC \cite{Apollinari:2015bam} is the foreseen upgrade of the LHC collider to achieve instantaneous luminosities a factor of five larger than the present LHC nominal value. The upper limit on BR$(Z\rightarrow\gamma \dph)$ at the HL-LHC has been derived from the one obtained for the LHC by taking into account the increase in luminosity \cite{Apollinari:2015bam}. 

\subsubsection{The ATLAS detector}

The ATLAS detector, used in the following as a benchmark experimental framework for simulations at hadronic colliders,  is a 46 m long cylinder, with a diameter of 25 m. 
It consists of six different cylindrical subsystems wrapped concentrically in layers around the collision point to record the trajectories, momenta, and energies of the particles produced in the collision final states.
A magnet system bends the paths of the charged particles so that their momenta can be measured.
The four major components of the ATLAS detector are the Inner Detector (ID), the Calorimeter, the Muon Spectrometer and the Magnet System. 
The ID components, embedded in a solenoidal 2 T magnetic field
cover a pseudorapidity range of $|\eta|< 2.5$.
%
The calorimeters cover the range $|\eta|<4.9$, using different techniques suited to the widely varying requirements of the physics processes of interest and of the radiation environment. 
The total thickness, including 1.3 $\lambda$ from the outer support, is 11 $\lambda$ at $\eta = 0$.
In the muon spectrometer,
 over the range $|\eta|< 1.4$, magnetic bending is provided by the large barrel toroid.  For $1.6<|\eta|<2.7$, muon tracks are bent by two smaller
end-cap magnets. 
Over the so-called transition region $1.4<|\eta|< 1.6$, magnetic deflection is provided by a combination of barrel and end-cap fields \cite{Aad:2008zzm}. 

\subsection{\label{sec:methods1}Methods}

\subsubsection{\label{sec:mc_lhc}Monte Carlo simulation}

At the LHC, we study the discovery potential for the decay $Z\rightarrow  \gamma \dph$  through the signal process $ p p \rightarrow Z  \rightarrow  \gamma \dph$. 
%
The $Z$-boson production fiducial cross section $\sigma_{fid}$ was calculated at the Leading Order (LO) in QCD using {\sc MadGraph5\textunderscore aMC@NLO} \cite{Alwall:2014hca} with parton distributions from NNPDF23 \cite{Ball:2012cx}, summing over all fermionic $p p \rightarrow Z/\gamma^* \rightarrow f  \bar{f}$ contributions and imposing on the $Z$ decays the same fiducial cuts required on the signal single detected photon, that is,  a minimum $p_T$ of 10 GeV and $|\eta|<2.5$.
%
The obtained $Z$-boson production fiducial cross-section turns out to be equal to 
\be
\sigma_{fid}= (2.504 \pm 0.006) \times 10^{4} \, \text{pb}\,,
\ee
with an expected number $N_Z$ of $Z$-bosons produced at the LHC at its design luminosity of $300 \, \text{fb} ^{-1}$ 
of $N_Z=7.5 \times 10^9$.
At the HL-LHC, at design luminosity of $3 \, \text{ab} ^{-1}$, the predicted 
$N_Z$ turns out to be $7.5 \times 10^{10}$, that is, ten times more.
%
%

The crucial ingredient in searching in proton collisions for the decay $Z\rightarrow  \gamma \dph$  is the separation of the signal from the background processes, which is in general very challenging.

The three processes taken into account as main background sources are 
\begin{subequations}
\begin{align}
  \label{eq:LHCbck1} p \, p &\rightarrow \gamma + \text{jets,}\\ 
   \label{eq:LHCbck2} p \, p &\rightarrow \gamma \, \nu \bar{\nu} \text{,\;\;\; and}\\
   \label{eq:LHCbck3} p \, p &\rightarrow e^+ \, \nu_e/e^- \, \bar{\nu}_e.
\end{align}
\end{subequations}
Regarding process (\ref{eq:LHCbck1}), the unreconstructed energy from jet clustering can mimic missing energy coming from a \dph. In process (\ref{eq:LHCbck2}), each neutrino $\nu$ has the same signature of a massless dark photon, appearing as missing momentum in the $Z \rightarrow \gamma + X$ final state. The same holds true for the neutrinos in the process (\ref{eq:LHCbck3}), where electrons are wrongly reconstructed as photons and pass the photon identification requirements \cite{Aaboud:2018yqu}.

Both signal and background events were generated at the leading order (LO) using {\sc MadGraph5\textunderscore aMC@NLO}, which allows to compute the full hard process matrix element, including spin correlations. The $Z\gamma\dph$ interaction was modeled adapting the {\sc UFO} \cite{Degrande:2011ua} from \cite{Degrande:2012wf}, which in our notation corresponds to assuming $C_i=C$ in \eq{LeffMD} by fixing three out of the nine $d_M^f$ couplings, thus effectively simulating a simplified version of the full Lagrangian in \eq{LeffMD} with one single free parameter $C$, whose value was fixed during simulation to get conventionally $\mbox{BR}(Z\rightarrow $\dph$ \gamma)=0.5$.

\begin{widetext}
\begin{center}
\begin{table*}[h!]
\begin{tabular}{c c c c c }
 \multicolumn{4}{c}{\bf }\\ 
 \hline
 Process  & Slice & $ N_\text{sim} $ & $\sigma$ (pb) &  Selection \\ \hline
$p \, p  \rightarrow \gamma \dph $ & -  & 150000 & $\quad (2.504 \pm 0.006) \times 10^{4} $ & - \\
\hline 
$p \, p   \rightarrow \gamma \nu \bar{\nu}$& -  & 150000 & $13.9 \pm 0.2 $ & - \\
\hline 
 $p \, p   \rightarrow \gamma+ $jets& I    & 14166722 & $(8.31 \pm 0.01)\times 10^{4} $  & - \\
\hline 
$p \, p   \rightarrow \gamma +$jets& II    & 281057 & $ 645 \pm 2 $ & $E_{\gamma}>$ 300 GeV  \\
\hline
$p \, p   \rightarrow \gamma +$jets& III    & 3860000 & $(2.468 \pm 0.005)\times 10^{4} $ & $p_T^{\gamma}>$ 30 GeV  \\
\hline
\end{tabular}
\caption{\small Simulated processes with the corresponding number of generated events ($N_\text{sim}$) and cross-sections. The signal cross-section is conventionally set following BR$(Z\rightarrow \gamma \dph)=0.5$. Selection cuts for the single hard photon required during generation are also shown, when relevant. The uncertainties quoted on cross-sections are purely statistical.}
\label{tablhc}
\end{table*}
\end{center}
\end{widetext}

Parton shower and hadronization effects were simulated using {\sc Pythia8} \cite{Sjostrand:2006za,Sjostrand:2014zea}, while the ATLAS detector response simulation was performed using {\sc Delphes} \cite{deFavereau:2013fsa}. 
The simulated processes and the corresponding number of generated events and cross-sections are shown in Table~\ref{tablhc}.  

%
In order to optimize the generation step, all samples were produced by requiring a single isolated photon with $p_T>10$ GeV and a minimum parton transverse momentum $p_T>20$ GeV in the hard process, where relevant.  
%
%

In the same fashion, three distinct samples were simulated for the process $p \, p   \rightarrow \gamma+ $jets, referred to in Table \ref{tablhc} and in the following as \emph{slices}: slice I was generated without a cut in the photon energy, slice II required a minimum photon energy $E_\gamma^\text{min}$ before detector smearing effects (``particle level'') of 300 GeV, while slice III required a minimum photon transverse momentum $p_{T,\gamma}^\text{min}$ of 30 GeV. The three slices were eventually merged into a single final sample, properly weighting by the corresponding cross sections  events in the overlapping regions. 

\subsubsection{\label{sec:evrec}Event reconstruction}

In order to model a detector as close as possible to ATLAS, the following conditions were specifically implemented.
Charged particles were assumed to propagate in a magnetic field of 3.8 T enclosed in a cylinder of radius 1.29 m and length 6 m.
Photons were reconstructed from clusters of simulated energy deposits in  the electromagnetic  calorimeter  measured  in projective towers with no matching tracks.  Photons were identified and isolated by requiring  the energy deposits in the calorimeters to be within a cone of size 
\be
\Delta R=\sqrt{\Delta \eta^2 + \Delta \phi^2} = 0.1
\ee
 around the cluster barycentre. Candidate  photons were required to have transverse energy $E_T>10$ GeV, in order to simulate ATLAS photon efficiency calibrations selections \cite{Aaboud:2018yqu}, and to be within $|\eta| \le 2.5$.

\begin{table}[h!]
\centering
\begin{tabular}{c}
 \multicolumn{1}{c}{\bf }\\ 
 \hline
 Cut  \\ \hline
 $p_T^{\gamma} >35 \, \text{GeV} $ \\
\hline 
 $E_T^\text{miss} > 40 \,\text{GeV} $ \\
\hline 
$\Delta\phi(\gamma,E_T^\text{miss}) > 2.8$ rad\\
\hline
$80 \,\text{GeV} < M_T <105 \,\text{GeV} $ \\
\hline 
\end{tabular}
\caption{\small Cuts applied to maximize $s/\sqrt{b}$\label{tabcuts}. When deriving limits, the cut on $M_T$ is applied only when using $E_\gamma$ as discriminating variable (see \ssecref{sec:statm} ).}
 \end{table}

Electrons and muons were reconstructed from clusters of energy deposits in the electromagnetic calorimeter matched to a track within $\Delta R = 0.5$, without any attempt to simulate independenlty the muon detector response.

Jets were reconstructed with the anti-$k_T$ algorithm \cite{Cacciari:2008gp} with a radius parameter $R = 0.5$ from clusters of energy deposits at the electromagnetic  scale in the calorimeters. This scale reconstructs the energy deposited by electrons and photons correctly but does not include any corrections for the loss of signal for hadrons due to the non-compensating character of the ATLAS calorimeters.   A correction
to calibrate the jet energy was then applied,
such that a sample of hadrons of a given energy is reconstructed with that same energy. Candidate jets were required to have $p_T > 20$ GeV. 

The missing transverse energy $E_{T}^\text{miss}$ was computed as the transverse component of the negative vector sum of the momenta of the candidate physics objects.

\subsubsection{\label{sec:evsel}Event selection}

\begin{table}[tb] 
\centering
\begin{tabular}{c c c }
 \multicolumn{3}{c}{\bf }\\ 
 \hline
Cut & $	\epsilon_{s}$ & $	\epsilon_{b}$  \\ \hline
Preselection & 0.49 & 0.24 \\ \hline
 $p_T^{\gamma} $ and $E_T^\text{miss}$ & 0.27 & $2.3 \times 10^{-4}$  \\
\hline 
$p_T^{\gamma} $, $E_T^\text{miss}$ and $M_T$ & 0.22 & $1.2 \times 10^{-4}$ \\
\hline 
$p_T^{\gamma} $, $E_T^\text{miss}$, $M_T$ and $\Delta\phi(\gamma,E_T^\text{miss})$ & 0.19 & $8.7 \times 10^{-5}$ \\
\hline
\end{tabular}
\caption{\small Cut efficiencies after preselection and selection at the LHC for both the signal $\epsilon_s$ and the sum of the considered backgrounds $\epsilon_b$.}
\label{tab:efflhc}
\end{table}


Only events with at least one reconstructed photon and no jets in the final state were selected, in order to improve discrimination against the main backgrounds. This requirement, together with the loose cuts applied at generation level, will be referred to in the following as ``preselection".

\begin{figure} [h!]
\begin{center}
 \label{fig:gammaElhc_a}\includegraphics[width=7cm]{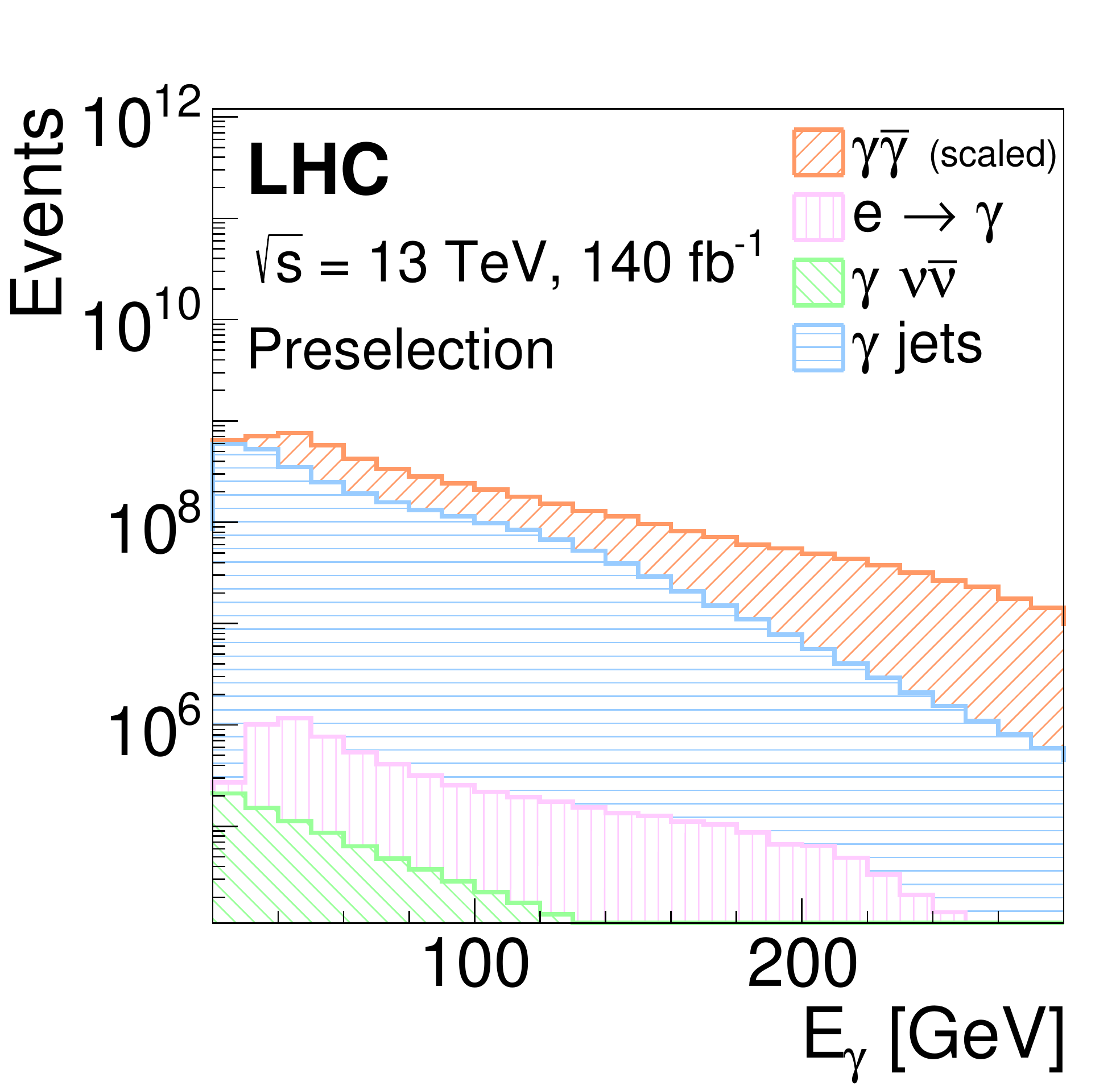}\\
 \label{fig:gammaElhc_b}\includegraphics[width=7cm]{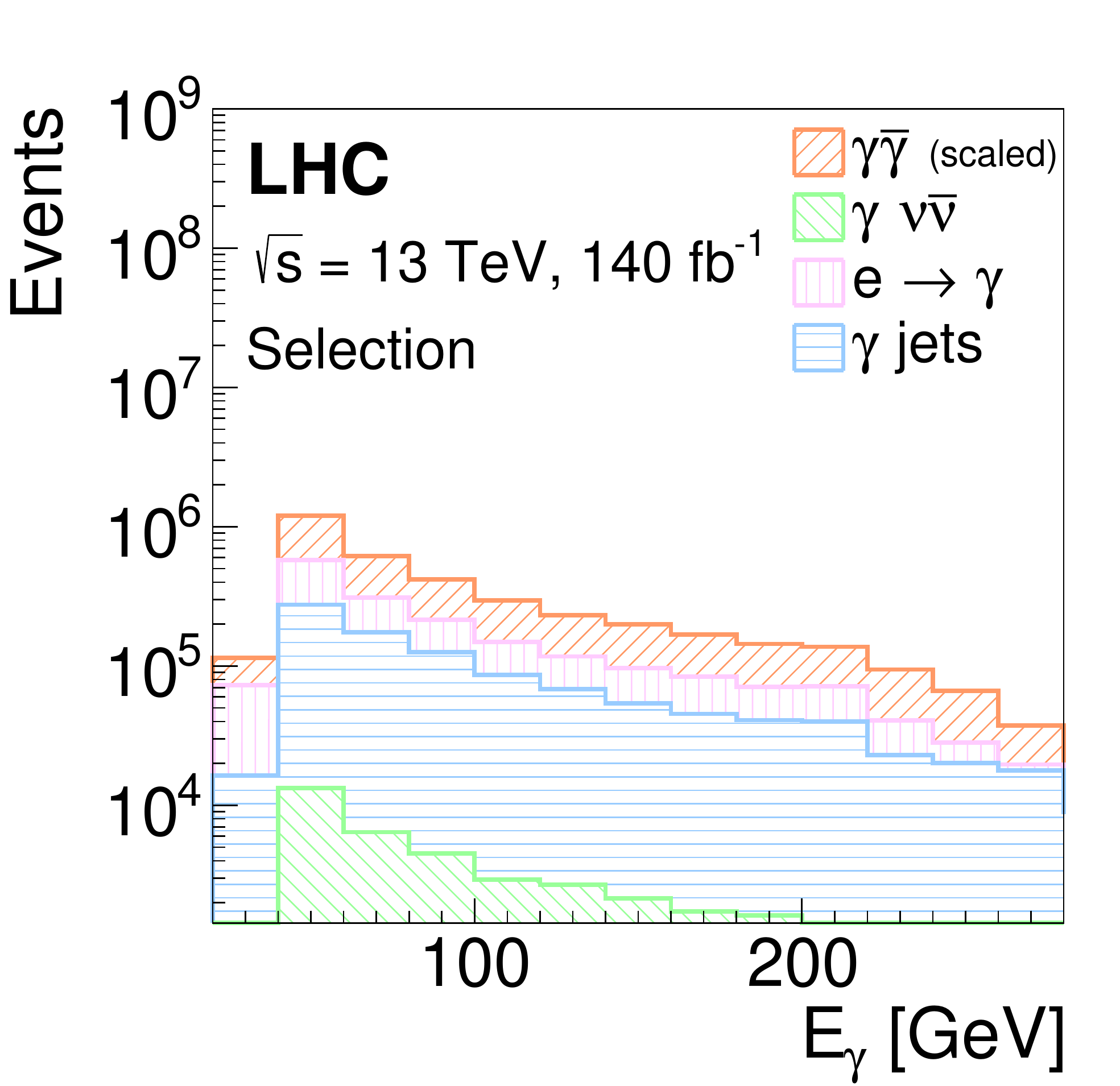}
 \caption{ \small Photon energy $E_{\gamma}$ distributions for the signal and background processes, passing (on the top) 
 preselection and (on the bottom)
 selection requirements. The signal distribution is normalized to the total estimated background yield. }
\label{fig:gammaElhc}
\end{center}
\end{figure}

\begin{figure} [h!]
\centering
\includegraphics[width=8.2cm]{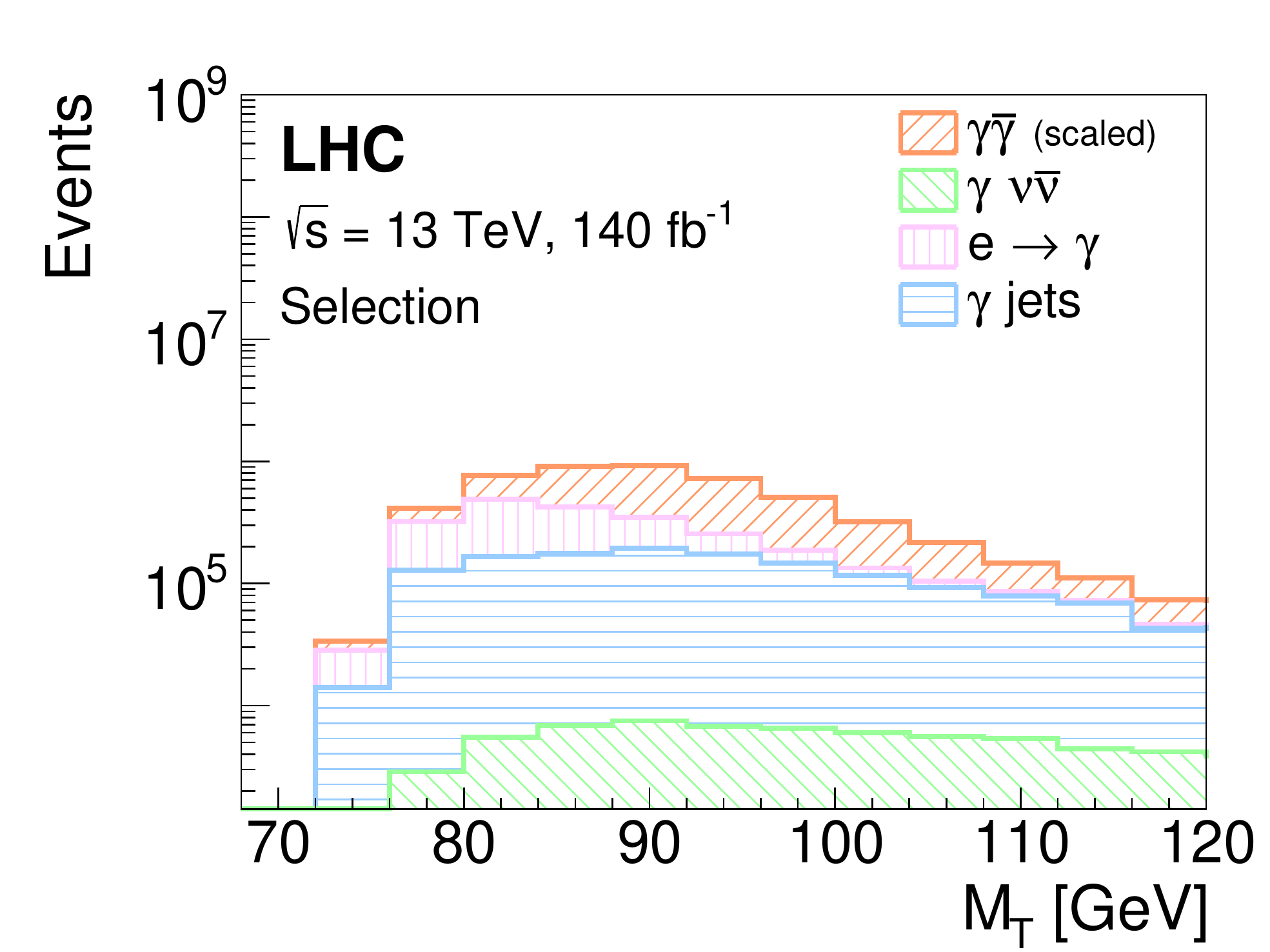}
\caption{\small Transverse mass $M_T$ distribution for the signal and background processes after selection cuts in $p_T^{\gamma}$, $E_T^\text{miss}$ and $\Delta\phi(\gamma,E_T^\text{miss})$. The signal distribution is normalized to the total estimated background yield.}
\label{fig:mtcuts}
\end{figure}

In order to maximize the sensitivity of the search we investigated the possible application of several kinematic cuts targeting an increase of the signal over square root of background yields ratio
$ s/\sqrt{b}$ (see Figures \ref{fig:gammaElhc} and \ref{fig:mtcuts}). The resulting cuts are summarized in Table \ref{tabcuts}, where the invariant transverse mass $M_T$ built with the photon and the missing transverse energy is defined through the formula
\be
M_T^2 = 2E_T^\text{miss}\cdot p_T^{\gamma} \cdot [1-\cos\Delta\phi(\gamma,E_T^\text{miss})] \, ,
\label{eq:mtdef}
\ee
where $p_T^{\gamma}$ in the transverse component of the photon momentum,
and $\Delta\phi(\gamma,E_T^\text{miss})$ is the angle between the photon
and the missing transverse momenta in the transverse plane.

In the following, the cuts reported in Table \ref{tabcuts} will be referred to as ``selection''.


%
The cut efficiencies due to preselection and selection are reported in Table \ref{tab:efflhc}.

\subsubsection{\label{sec:statm}Statistical Methods}

 We use a simple binned likelihood function ${\cal L}(\mu,\boldsymbol{\theta})$ constructed as a product of Poisson probability terms over all bins considered~\cite{ATLAS:2019oik}. Either $M_T$ or $E_\gamma$ are used as discriminating variable, whichever gives the  better limits. 

The likelihood function is implemented in the \textsc{RooStats} package~\cite{Moneta:2010pm}.
It depends on the signal-strength parameter $\mu$, a multiplicative factor 
that scales the number of expected signal events,
and $\boldsymbol{\theta}$, a set of nuisance parameters (NPs) that encode the effect of systematic uncertainties on the background expectations, which are implemented in the likelihood function as Gaussian constraints. One should notice that, in our conventions, we can identify the parameter $\mu$ with the $\text{BR}(Z\rightarrow \gamma \dph)$ in the limit $\mu \ll 1$, which always holds throughout the paper when deriving limits.
Individual sources of systematic uncertainty are considered to be uncorrelated. The statistical uncertainty of the MC events is not taken into account in ${\cal L}(\mu,\boldsymbol{\theta})$, while increasing statistics of MC samples in specific regions of phase space when needed (see e.g. \ssecref{sec:mc_lhc}).

The test statistic $q_\mu$ is defined as the profile likelihood ratio
\be
q_\mu = -2\ln \left( \frac{{\cal L}(\mu,\boldsymbol{\hat{\hat{\theta}}}_\mu)}{{\cal L}(\hat{\mu},\boldsymbol{\hat{\theta}})} \right)\, ,
\ee
 where $\hat{\mu}$ and $\boldsymbol{\hat{\theta}}$ are the values of the parameters that maximize the likelihood function (with the constraint $0\leq \hat{\mu} \leq \mu$), and $\boldsymbol{\hat{\hat{\theta}}}_\mu$ are the values of the NPs that maximize the likelihood function for a given value of $\mu$. The test statistic $q_\mu$ is implemented in the \textsc{RooFit} package~\cite{Verkerke:2003ir}. Assuming the absence of any significant excess above the background expectation, upper limits on $\text{BR}(Z\rightarrow \gamma \dph)$ are derived by using $q_\mu$ and the CL$_{\textrm{s}}$ method~\cite{junk:1999kv,read:2002hq}. Values of $\text{BR}(Z\rightarrow \gamma \dph)$ (parameterized by $\mu$) yielding CL$_{\textrm{s}}$ $<0.05$, where CL$_{\textrm{s}}$ is computed using the asymptotic approximation~\cite{cowan:2010js}, are excluded at $95$\% CL.

\subsection{\label{sec:res1}Results}

We can  now derive the upper limit on the BR($Z  \rightarrow \gamma \dph$)  attainable at the LHC. 

Let us first look at the best case scenario in which systematic uncertainties are negligible.  We find
\be
\mbox{BR} (Z  \rightarrow \gamma \dph)=  4.0 \times 10^{-6}\, . 
\label{eq:brlhcc0}
\ee

Though the BR in \eq{eq:brlhcc0} might compete with the LEP result $\mbox{BR} (Z  \rightarrow \gamma \dph) < 10^{-6}$ \cite{Abreu:1996vd} after taking into account the combination of the ATLAS and CMS experiments, or assuming HL-LHC luminosities, this result is  weakened by the effect of systematic uncertainties which are unavoidable in the real case.

\begin{table}[h]
\centering
\begin{tabular}{c c c }
 \multicolumn{3}{c}{\bf }\\ 
 \hline
 Observable $\quad$ & Systematic uncertainty $\quad$ & $c_i$ \\ \hline
  $E_T^\text{miss}$ & 1$\% $ & $  0.04  $ \\
\hline 
$p_T^{\gamma}$  & 0.3 $\%$ & $<0.01$ \\
\hline 
$\sigma_{b}$ & 3.7$\%$ & 0.04 \\
\hline 
$F_{e\rightarrow\gamma}$ & 70$\%$ & - \\
\hline 
\end{tabular}
\caption{\small Main systematic uncertainty sources on the background yields and corresponding overall relative impact. The sources of systematic uncertainties were taken from \cite{Aaboud:2017cbm}. }
\label{tabc}
\end{table}

The level of uncertainty on the background yields will depend on several factors, possibly decreasing with the life of the accelerator due to welcome efforts of the collaborations in understanding and constraining systematic effects. Here an estimate of the overall relative impact $\displaystyle{c_i = {\Delta b_i }/{b}}$ of the  systematic $i$  on the total background yields is attempted, by analysing a subset of possible systematic uncertainty sources chosen, among the set affecting a search with analogue signature \cite{Aaboud:2017cbm}, as the three with highest impact: the uncertainty on $E_T^\text{miss}$, the jet energy scale uncertainty and the theoretical uncertainty on the  modeling of $\sigma_{b}$.
In order to do this, we varied up and down the source of the uncertainty by one standard deviation, taken from \cite{Aaboud:2017cbm}: we eventually computed the resulting value of $c_i$ using the highest variation $\Delta b_i=\max{\left\lbrace \Delta b_i^\text{up},\Delta b_i^\text{down} \right\rbrace}$, as summarized in Table  \ref{tabc}. A dedicated uncertainty on $F_{e\rightarrow \gamma}$, the $e\rightarrow \gamma$ fake-rate, was assigned to the background from electrons wrongly reconstructed and misidentified as photon candidates \cite{Aaboud:2018yqu}. This systematic uncertainty conservatively covers both converted and unconverted photon fake rates in the central region.

 
After properly assigning a NP to each source of systematic uncertainty  from Table  \ref{tabc} in the likelihood function of \ssecref{sec:statm}, we compute the 95\% CL upper limit on $\mbox{BR}(Z  \rightarrow \gamma \dph)$ considering $L=140\text{ fb}^{-1}$ to be 
\begin{equation} \label{eq:LHCresult}
\mbox{BR}(Z  \rightarrow \gamma \dph) < 8.0 \times 10^{-6}.
\end{equation}

\begin{figure} [h!]
\centering
\includegraphics[width=8.2cm]{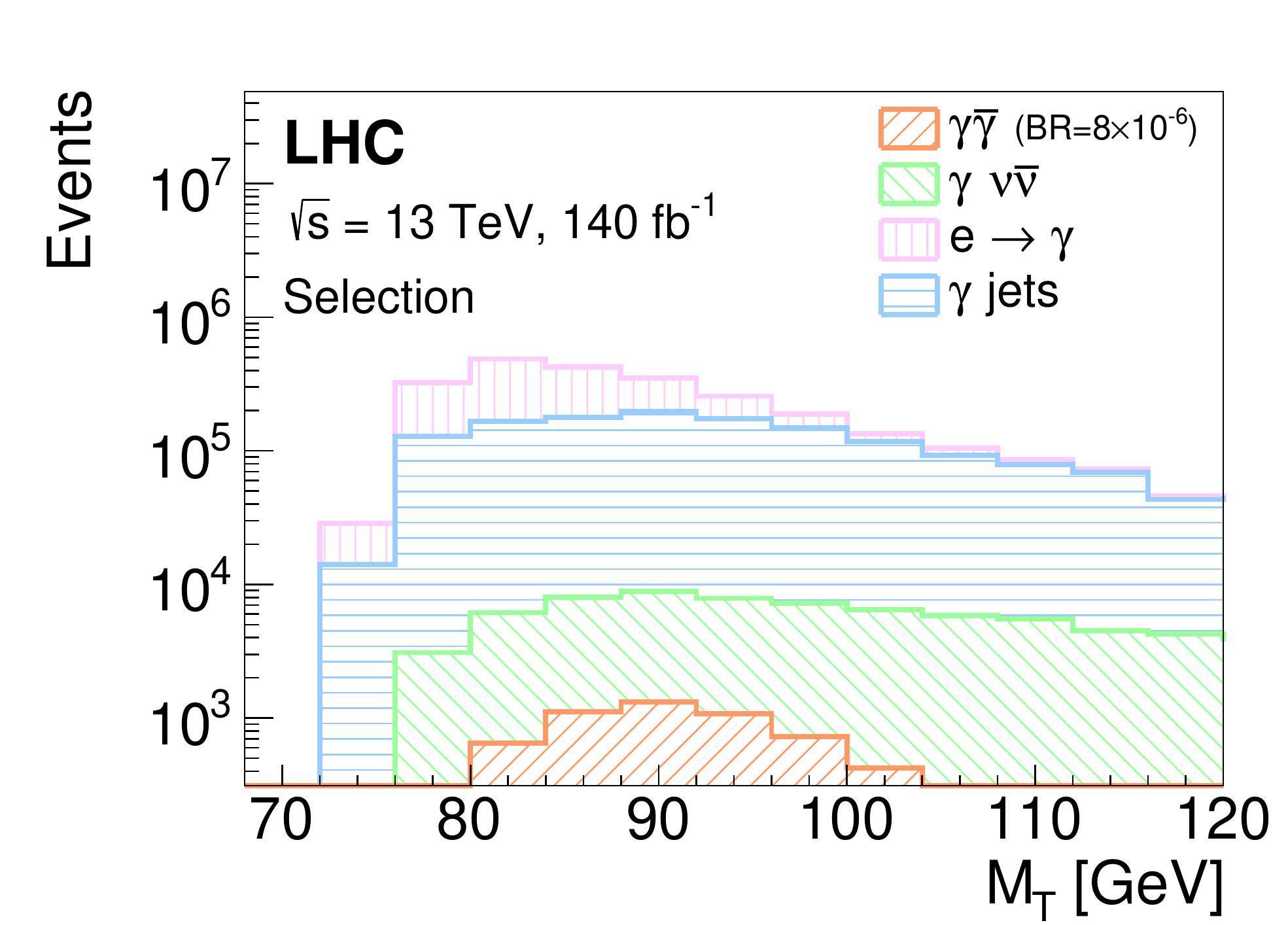}
\caption{\small Transverse mass $M_T$ distribution for the signal and background processes after selection cuts in $p_T^{\gamma}$, $E_T^\text{miss}$ and $\Delta\phi(\gamma,E_T^\text{miss})$ and by scaling the signal distribution by the LHC upper limit $\mbox{BR}(Z  \rightarrow \gamma \dph)= 8 \times 10^{-6}$.}
\label{fig:mtcutsstackBR}
\end{figure}

The $M_T$ distribution after selection cuts, with signal yields normalized according to \eq{eq:LHCresult}, is shown in Figure~\ref{fig:mtcutsstackBR}. 
This result, compared with \eq{eq:brlhcc0}, highlights how the study and improvement in the control of systematic uncertainties will be a key feature at the LHC.
%
The corresponding HL-LHC  upper limit on $\mbox{BR}(Z  \rightarrow \gamma \dph)$ can be estimated under the  assumption that the systematic uncertainties will decrease by a factor $1/\sqrt{L} $. Therefore the upper limit on $\mbox{BR}(Z  \rightarrow \gamma \dph)$ in \eq{eq:LHCresult} can simply be multiplied by a factor $\sqrt{L_\text{LHC}/L_\text{HL-LHC}}= \sqrt{140 \text{ fb}^{-1}/3000 \text{ fb}^{-1}}$,  obtaining 
\be
 \mbox{BR}(Z  \rightarrow \gamma \dph) < 2 \times 10^{-6}\, ,
\ee
which represents the estimate for the best upper limit reachable by a single experiment at the HL-LHC. 

From this discussion, it is clear that a search program for the process $p p \rightarrow Z \rightarrow \gamma \dph$ will hardly compete with the LEP result, mainly  because of
 the large background contamination. This negative result does not come unexpected but the exercise of this section is still useful in showing quantitatively the kind of problems one encounters in trying to study this particular process at a hadron collider.
 
More promising results can be obtained at future lepton colliders, to which we now turn.

\section{\label{sec:Future Machines}Future Lepton colliders}

Two circular lepton colliders have currently been proposed. The first one is part of 
the Future Circular Collider project (FCC), whose integrated program foresees operations in two stages: initially an electron-positron collider (FCC-ee) serving as a Higgs and electroweak factory running at different CM energies, followed by a proton-proton collider at a collision energy of 100 TeV.

The FCC-ee \cite{FCC-ee:2020} is a high-luminosity, high-precision, 100 km circumference storage ring collider, designed to provide $e^+e^-$ collisions with centre-of-mass energies from 88 to 365 GeV. The CM operating points with most physics interest are around 91 GeV ($Z$-boson pole), 160 GeV ($W^{\pm}$ pair-production threshold), 240 GeV ($ZH$ production) and 340-365 GeV ($t \bar t$ threshold and above). The machine should  deliver peak luminosities above 10$^{34}$cm$^{-2}$s$^{-1}$ per experiment at the $t \bar t$ threshold and the highest ever luminosities at lower energies, with an expected total integrated luminosity $L=150\text{ ab}^{-1}$ at the $Z$ pole. 

The other planned electron-positron machine is the Circular Electron Positron Collider (CEPC \cite{CEPCStudyGroup:2018rmc}), which  is expected to collide electrons and positrons at different CM energies: 91.2 GeV, 160 GeV and 240 GeV \cite{CEPCStudyGroup:2018ghi}. This machine is expected to provide a total integrated luminosity $L=16\text{ ab}^{-1}$ at the $Z$ pole.

These new lepton colliders will  produce huge statistics samples of events (of the order of Tera  $Z$ bosons), allowing many measurements with unprecedented accuracy, and the discovery and study of rare $Z$, $W$, Higgs boson and top decays. Besides offering the ultimate investigations of electroweak symmetry breaking, these precision measurements will be highly sensitive to the possible existence of yet unknown particles, with masses up to about 100 TeV. Sensitive searches for particles with couplings much smaller than weak, such as sterile neutrinos, can be envisioned as well. 

  We look into the feasibility of a search for a massless dark photon at these new machines, focusing on the center of mass energy of 91.2 GeV, which represents the most promising setup for the process here considered.
  

\subsubsection{An FCC-ee detector: IDEA}

The IDEA detector concept~\cite{FCC-ee:2020}, developed specifically for FCC-ee, is based on established technologies resulting from years of R$\&$D. Additional work is, however, needed to finalise and optimise the design. 

The detector comprises a silicon pixel vertex detector, a large-volume extremely-light short-drift wire chamber surrounded by a layer of silicon micro-strip detectors, a thin, low-mass superconducting solenoid coil, a pre-shower detector, a dual-readout calorimeter, and muon chambers within the magnet return yoke. Electrons and muons with momenta above 2 GeV and unconverted photons with
energies above 2 GeV can be identified with efficiencies of nearly 100$\%$ and with negligible backgrounds. The photon energy should be measured to a precision of $11\%/ \sqrt{E} \oplus  1\%$.

\subsubsection{\label{sec:cepc}The CEPC detector}

Two primary detector concepts have been developed for the construction of the CEPC detector: a baseline, with two approaches for the tracking systems, and an alternative one, with a different strategy for meeting the jet resolution requirements. In the following we briefly describe the baseline approach, being the one used for simulations. 

The baseline detector concept~\cite{CEPCStudyGroup:2018ghi} incorporates the particle flow principle  with a precision vertex detector, a Time Projection Chamber (TPC) and a silicon tracker, a high granularity calorimetry system, a 3 Tesla superconducting solenoid followed by a muon detector embedded in a flux return yoke. In addition, five pairs of silicon tracking disks are placed in the forward regions at either side of the Interaction Point (IP) to enlarge the tracking acceptance from $|\cos \theta | <0.99$ to $|\cos \theta | <0.996$.


The performance of the CEPC baseline detector concept have been investigated with full simulation, as summarized in the following with focus solely on features which have some impact in the study we are presenting here. Electrons and muons with momenta above 2 GeV and unconverted photons with
energies above 5 GeV can be identified with efficiencies of nearly 100$\%$ and with negligible backgrounds. 
The photon energy should be measured to a precision better than $20\%/ \sqrt{E} \oplus  1\%$.
The ionization energy loss ($dE/dx$) will be measured in the TPC, allowing the identification of low momentum charged particles.
Combining the measurements from the silicon tracking system and the TPC, the track
momentum resolution will reach $\Delta(1/p_T) \sim 2 \times 10^{-5} \, \text{GeV}^{-1}$. 

\subsection{\label{sec:methods2} Methods}

\subsubsection{Monte Carlo simulation samples}


The   fiducial cross section for the SM $Z$-boson production 
at lepton colliders with a CM energy of 91.2 GeV and, following the same procedure described in \ssecref{sec:mc_lhc}, 
 with decays within $|\eta|<3$ is 
 \be
\sigma( e^+\,e^-\,\rightarrow Z/\gamma^*\rightarrow f \bar{f} ) = (6.19 \pm 0.01) \times 10^{4} \; \mbox{pb}\, .
\ee
%

The signal process $e^+ e^-  \rightarrow  Z \rightarrow \gamma \dph$   has a distinctive experimental signature. Both the photon and the massless dark photon are monochromatic with an energy of $M_Z/2$ at the $Z$-factory. The massless dark photon  has a neutrino-like signature, appearing as missing momentum in the $Z \rightarrow \gamma + X^0$ final state, in association with a peak of photon events around the mentioned energy values.

Two main processes can contribute to background events: 
\begin{subequations}
\begin{align}
  \label{eq:LEPCOLLbck1} e^+ \, e^- &\rightarrow \gamma \, \nu \bar{\nu}\text{, \;\;\; and}\\ 
   \label{eq:LEPCOLLbck2} e^+ \, e^- &\rightarrow \gamma \, e^+ \, e^-.
\end{align}
\end{subequations}

 In process (\ref{eq:LEPCOLLbck1}) the photon is the result of initial state radiation by either the electron or the positron, and the $\nu \bar{\nu}$ pair is produced either by the decay of a $Z$-boson produced in the $s$-channel or by $W$-exchange in the $t$-channel.
The radiative Bhabha reaction of process (\ref{eq:LEPCOLLbck2}) contribute to background events when both the final state electron and positron escape detection. However the number of background events from this process strongly depends on the geometry of the detector and on the presence of un-instrumented regions, as observed at LEP \cite{Abreu:1996vd} and shown in the following sections for the CEPC and FCC-ee detectors. 

The hard process, parton-shower and hadronization steps were simulated closely following the lines of \ssecref{sec:mc_lhc}. We checked explicitly that relevant kinematic distributions from reconstruction of events with the two different detector configurations of FCC-ee and CEPC closely match in event yields, up to Monte Carlo statistical fluctuations, the minor differences coming only from the slightly different energy resolutions. If not stated explicitly, in the following the configuration for the baseline CEPC detector concept must be understood, with results from the two detectors differing only by the corresponding integrated luminosity.

\begin{table*}[t!] 
\centering
\begin{tabular}{ c c c c c }
 \multicolumn{4}{c}{\bf }\\ 
 \hline
 Process & Slice & $ \quad \quad N_\text{sim}\quad  $ & $\sigma$  (pb) & $\quad E_{\gamma}^\text{min}$ (GeV) \\ \hline
$e^+ e^-  \rightarrow \gamma \dph$ & -  & 50000 & $(6.19 \pm 0.01) \times 10^{4}  $ & - \\
\hline 
$e^+ e^-  \rightarrow \gamma \nu \bar{\nu}$ & I & 5000000 & $5025.0 \pm 4.5 $ & -  \\
\hline 
$e^+ e^-  \rightarrow \gamma \nu \bar{\nu}$ & II & 500000 & $0.1599 \pm 0.0002  $ & 18 \\
\hline 
 $e^+ e^-  \rightarrow \gamma e^+ e^- $ & I  & 5000000 & $8100 \pm 1176$ & - \\
 \hline
 $e^+ e^-  \rightarrow \gamma e^+ e^- $ & II  & 500000 & $220.9 \pm 0.4 $ & 30  \\
\hline 
\end{tabular}
\caption{\small Simulated processes with the corresponding number of generated events ($N_\text{sim}$) and cross-sections. As in Table \ref{tablhc}, the signal cross-section is conventionally set following BR$(Z\rightarrow \gamma \dph)=0.5$. See  \ssecref{sec:mc_lhc} for the definition of the slices I and II. }
\label{tab222}
\end{table*}

The simulated processes and the corresponding number of generated events and cross-sections assuming collisions between positron and electron beams with $E_\text{beam}=$45.6 GeV are reported in Table \ref{tab222}. For each background process, two slices have been simulated with different $E_\gamma^\text{min}$ during generation, respectively of 18 and 30 GeV for  the $e^+ e^-  \rightarrow \gamma \nu \bar{\nu}$ and  $e^+ e^-  \rightarrow \gamma e^+ e^- $  process (see \ssecref{sec:mc_lhc}). 

\subsubsection{Event reconstruction} 

Charged particles were assumed to propagate in a magnetic field of 3.5 T homogeneously filling a cylindrical region of radius 1.81 m and length of 4.70 m.

Photons were reconstructed from clusters of energy deposits in  the  electromagnetic  calorimeter  measured in projective towers with no matching tracks. Photons were identified and isolated by requiring  the energy deposits in the calorimeters to be within a cone of size 
\be
\Delta R=\sqrt{\Delta \eta^2 + \Delta \phi^2} = 0.5
\ee
 around the cluster barycenter. Candidate  photons were required to have $E > 2$ GeV and to be within $|\cos \theta |\le 0.987$. 

Electrons and muons were reconstructed from clusters in the electromagnetic calorimeter with a matching a track. The criteria for their identification were similar to those used for photons.

Jets were reconstructed with the anti-$k_T$ algorithm  with a radius parameter $R = 0.5$ from clusters of energy deposits in the calorimeters (hadronic and electromagnetic).  Candidate jets were required to have $p_T >$20 GeV.

The missing energy vector $\Vec{p}_\text{miss}$  was computed as the negative vector sum of the momenta of the candidate physics objects.


\subsubsection{Event selection}

%
\begin{figure} [h!]
\centering
\includegraphics[width=8.2cm]{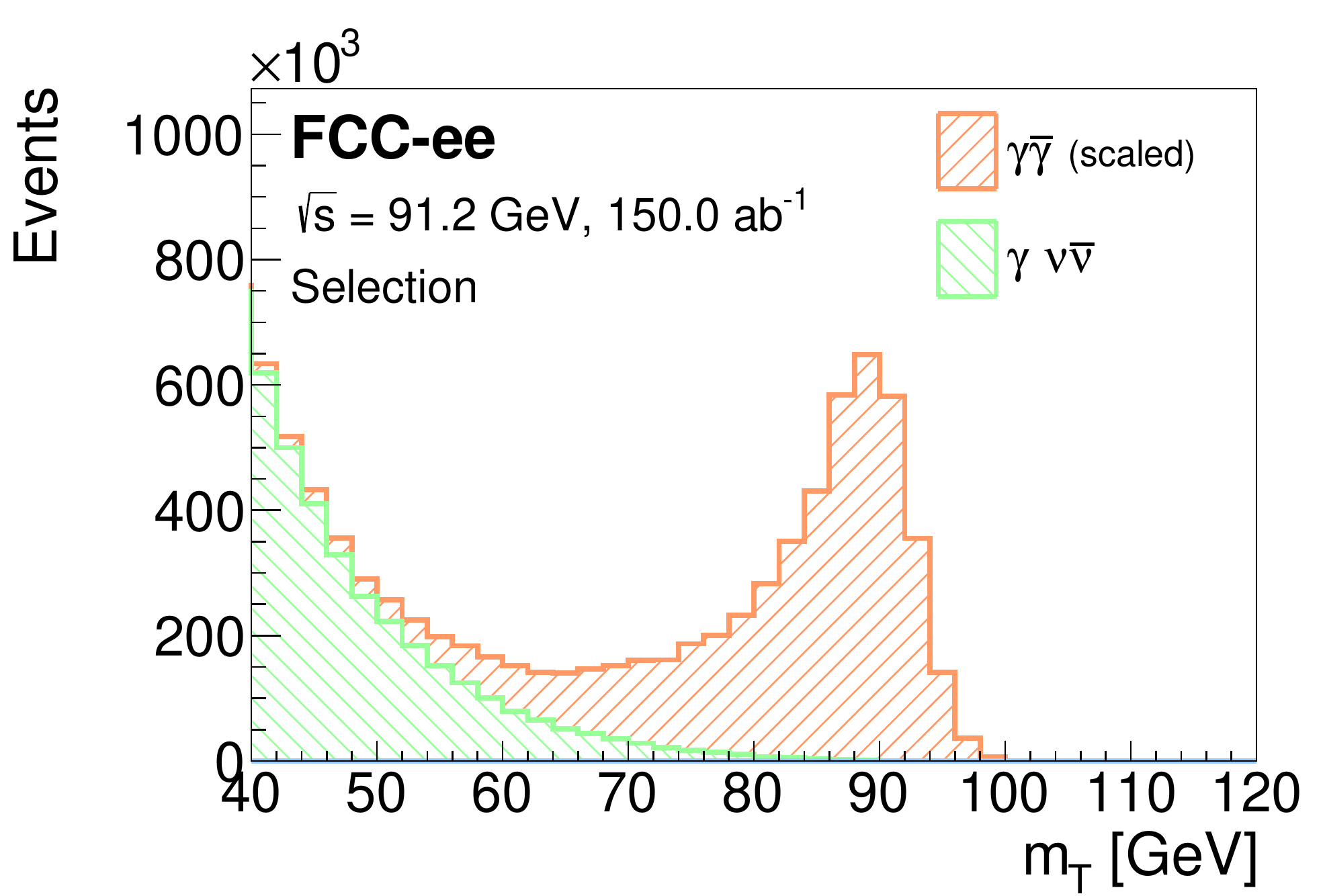}
\caption{$M_T$ distributions for the signal $e^{+} \, e^{-}  \rightarrow \gamma \dph$ (red) and $e^{+} \, e^{-} \rightarrow \gamma \nu \bar{\nu}$  (green). No event from the process $e^{+} \, e^{-}  \rightarrow \gamma e^{+}  e^{-} $ passed selection requirements. The signal distribution is normalized to the total estimated background yield.}
\label{fig:mtcepccutsZ}
\end{figure}

\begin{figure} [h!]
\centering
\includegraphics[width=8.2cm]{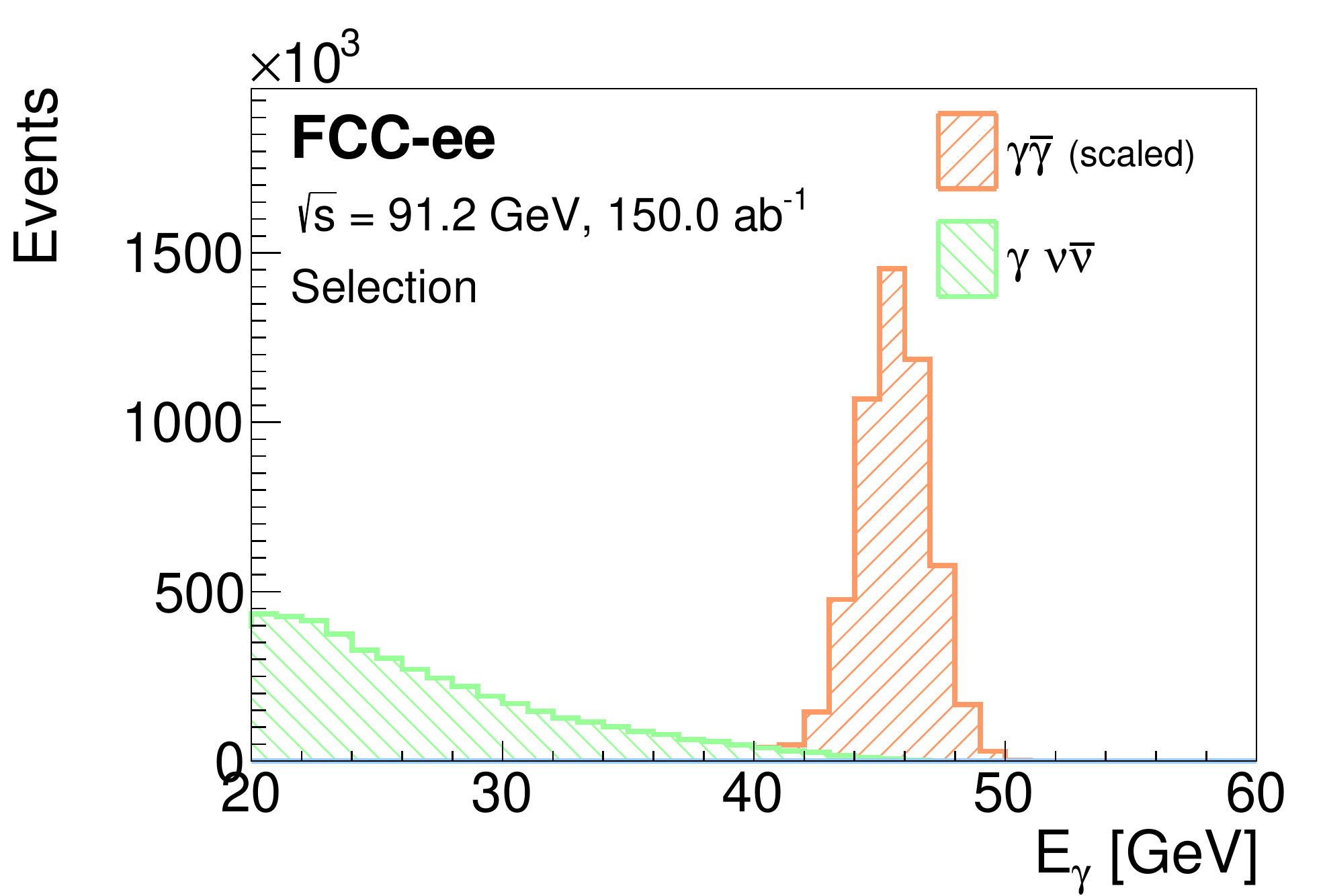}
\caption{ \small Photon energy $E_{\gamma}$ distribution after the cut in $\eta$ relative to the signal ($e^+ e^-  \rightarrow \gamma \dph$) and to the background processes ($e^+ e^-  \rightarrow \gamma \nu \bar{\nu}$ and $e^+ e^-  \rightarrow \gamma e^+ e^- $ ).The signal distribution is normalized to the total estimated background yield.}
\label{fig:gammaEcutsZ}
\end{figure}

Events were initially selected to have at least one photon and no charged particles in the final state. This requirement, together with the loose cuts applied at generation level, will be referred to in the following as ``preselection''.
Selection cuts were then applied to increase  the ratio $\displaystyle{{s}/{\sqrt{b}}}$ and improve the upper limit on the BR($e^+ e^-  \rightarrow \gamma \dph$), as summarized in Table \ref{tab:effcepc91}.

 At leptonic colliders the initial state of the process is fully determined (apart from initial-state-radiation effects) by the beam parameters, contrary to the hadron collider case, and the center of mass frame coincides with the laboratory frame. Yet we find instructive, in order to better understand the results of \secref{sec:res1}, to keep track of the same kinematical observables defined therein. The transverse invariant mass $M_T$ in events with one single photon and missing energy simplifies at particle level to the expression
\be
  M_T = 2 p_T^\gamma.
\ee
The $M_T$ and photon energy $E_\gamma$ distributions after selection are shown in Figures \ref{fig:mtcepccutsZ} and \ref{fig:gammaEcutsZ}. 
The preselection and selection efficiencies are reported in Table \ref{tab:effcepc91}: no event from the process $e^{+} \, e^{-}  \rightarrow \gamma e^{+}  e^{-} $ passed selection requirements, as expected from kinematic arguments, thanks to the absence of uninstrumented regions in both CEPC and FCC-ee detector designs.

\begin{table}[h!] 
\centering
\begin{tabular}{c c c }
 \multicolumn{3}{c}{\bf }\\ 
 \hline
Cut & $	\epsilon_{s} $ & $\epsilon_{b}$  \\ \hline
Preselection & $\quad$ 0.96 & $\quad$ 0.067 \\ \hline
 $\vert \cos \theta_{\gamma}\vert <$ 0.905 \quad & \quad 0.95 & $\quad 3.4 \times 10^{-3}$ \\
\hline 
 $\vert \cos \theta_{\gamma}\vert <$ 0.905 and $p_T^\gamma>$ 18 GeV & \quad 0.95 & $ \quad 2.3 \times 10^{-6}$ \\
\hline 
\end{tabular}
\caption{\small Preselection and selection efficiencies at the CEPC and FCC-ee detectors at 91.2 GeV. }
\label{tab:effcepc91}
\end{table}


\subsection{\label{sec:resLEPcoll}Results}


A lepton collider working at a a CM energy of $91.2$ GeV  is an actual $Z$-factory and the ideal place where to look for the small values of    the BR($Z  \rightarrow \gamma \dph$) we are after.  

Let us consider first the case of no systematic uncertainties.  The lepton colliders perform very well. The CEPC, with a total integrated luminosity $L=16\text{ ab}^{-1}$ yields
\be
\mbox{BR} (Z  \rightarrow \gamma \dph)=  7.1 \times 10^{-11}\, . 
\label{eq:brCEPC0}
\ee
The FCC-ee, with an expected total integrated luminosity of $L=150\text{ ab}^{-1}$, gives
\be
\mbox{BR} (Z  \rightarrow \gamma \dph)=  2.3 \times 10^{-11}\, . 
\label{eq:brFCCee}
\ee

The results in \eq{eq:brCEPC0} and \eq{eq:brFCCee} are not modified by taking into account the uncertainties on both the $\sigma_Z^{fid}$ ($0.01  \times 10^{4} $ pb) and  the luminosity $\Delta L/L = 10^{-4}$  \cite{CEPCStudyGroup:2018ghi}. Systematic effects on the upper limit on BR($e^+ e^-  \rightarrow \gamma \dph$) are found to be negligible.

The results summarized in Table \ref{tabconcl} show that, at both the CEPC and FCC-ee lepton colliders, running at the $Z$ pole mass, the limit on the BR($e^+ e^-  \rightarrow \gamma \dph$) will significantly improve the present LEP bound. 

\section{\label{sec:spin}Spin analysis}

Having discussed the discovery potential of a dark photon produced in association with a photon in $Z$-boson decays, in this last section we investigate how to establish the spin  of such a new neutral state.
Since nothing prevents the $Z$ boson to  decay into a photon and a \emph{pseudoscalar} massless neutral particle \dps, the latter can  be used as test hypothesis against the $J^P=1^-$ nature of the dark photon. 

No attempt is made here to optimize the search strategy for a pseudoscalar signal $Z\rightarrow a\gamma$, meaning that results presented in previous sections do not necessarily apply to this test hypothesis. 

\subsection{Methods}

\begin{figure}
\includegraphics[width=0.4\textwidth]{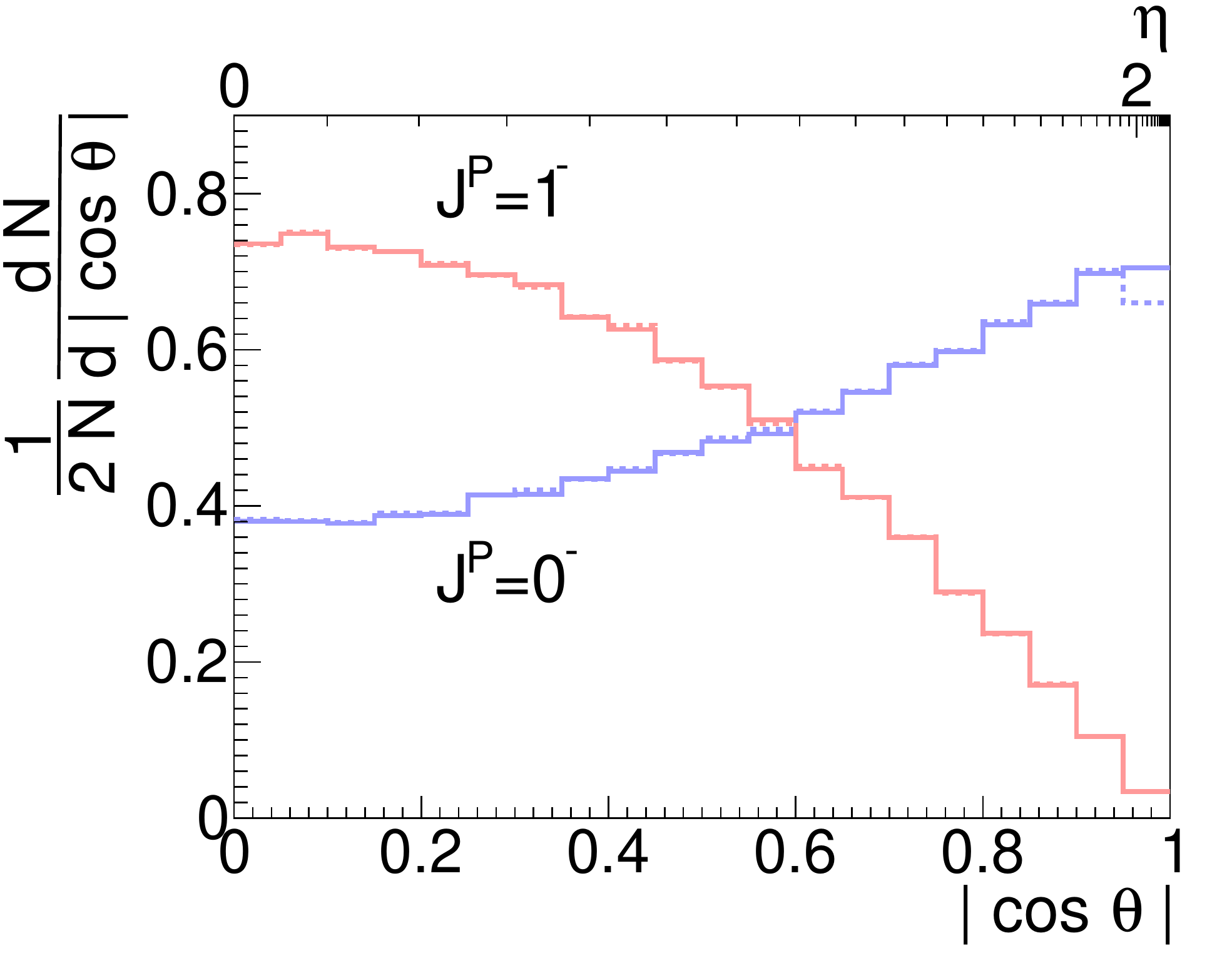}
\caption{\small Differential cross section as a function of the cosine of the detected photon polar angle $\theta$ when produced in association with a pseudoscalar (blue) or vector (red) massless dark particle, after background subtraction. Dashed lines describe the corresponding distribution when including detector smearing effects. The upper axis maps the same range in photon pseudorapidities.}
\label{fig:cosine}
\end{figure}

We now assume that the discovery of $\dph$ in $Z$-boson decays has  occured at a future $e^+e^-$ collider with $\sqrt{s}=M_Z$.
In this scenario, a good observable discriminating between the two $J^P=1^-,~0^-$ hypotheses is the cosine of the polar angle $\theta$ of the  detected photon~\cite{Comelato:2020cyk}. The corresponding distributions (Figure \ref{fig:cosine}) can be produced using the linear realization of the model \cite{Brivio:2017ije}, where the two relevant dimension-five operators regarding $\gamma\dps$  final states at the $Z$ peak  are
\bea
{\cal O}_{\dps Z \gamma} (x )&=\dps &Z_{\mu\nu}\tilde{A}^{\mu\nu}   \, ,\\
{\cal O}_{\dps \gamma \gamma} (x )&=\dps &A_{\mu\nu}\tilde{A}^{\mu\nu}   \, ,
\eea
with the latter contributing through interference with the first. We have checked that the independent contributions from 
${\cal O}_{\dps Z \gamma}$, ${\cal O}_{\dps \gamma \gamma}$ and their interference in the $\cos \theta$ distribution are indistinguishable in shape, as expected. One can then  apply  the following analysis to any combination of the $c_{\tilde{B}}$ and $c_{\tilde{W}}$ couplings, and in particular for the common and phenomenologically appealing $c_{\tilde{B}}=-\tan \theta_{W} c_{\tilde{B}}$ choice \cite{Brivio:2017ije}, which corresponds to set $g_{a\gamma\gamma}=0$.

Following the statistical treatment described in \ssecref{sec:statm}, a likelihood function $\mathcal{L}(J^{P},\mu,\boldsymbol{\theta})$ that depends on the spin-parity assumption  of  the  signal  is  constructed  as  a  product  of conditional probabilities over the binned distribution of the discriminating observable  $\cos \theta$:
\be
    \label{eq:likelihoodSpin}
    \mathcal{L}(J^{P},\mu,\boldsymbol{\theta}) = \prod_i^{N_{bins}} P\left(N_i \vert \mu S_i^{(J^P)}(\boldsymbol{\theta})+ B_i(\boldsymbol{\theta})\right) 
     \times \mathcal{A}(\boldsymbol{\theta}),
\ee
where, given the clean lepton collider environment, a good starting approximation is to assume the best case scenario in which all nuisance parameters $\boldsymbol{\theta}$ are sufficiently constrained by auxiliary measurements through the functions $\mathcal{A}(\boldsymbol{\theta})$, such to allow to safely neglect the contribution of systematic uncertainties on the background, which we assume subtracted from total yields with dedicated methods (see e.g. \cite{Pivk:2004ty}).  

Closely following \cite{Aad:2013120}, a proper test statistic $q$ is then chosen to be the logarithm of the likelihood ratio
\begin{equation}
\label{eq:teststatSpin}
 q= \log \dfrac{\mathcal{L}(J^{P}=1^{-},\hat{\hat{\mu}}_{1^{-}},\hat{\hat{\boldsymbol{\theta}}}_{1^{-}})}{\mathcal{L}(J^{P}=0^{-},\hat{\hat{\mu}}_{0^{-}},\hat{\hat{\boldsymbol{\theta}}}_{0^{-}})}.   
\end{equation}
In this case, given the rather simple form of both \eq{eq:likelihoodSpin} and \eqref{eq:teststatSpin}, they have been implemented directly in a \textsc{ROOT} \cite{Brun:1997pa} script. The distributions of the test statistic $q$ for both signals  
shown in Figure \ref{fig:llr} were obtained, as an example, using $n_{toys}=160000$ pseudo-experiments and with the specific choice of  $N=10$ signal events.

The distributions of $q$ are used to determine the corresponding $p_0$ values $p_0(J^{P}=1^{-})$ and $p_0(J^{P}=0^{-})$. For instance, the tested hypothesis $p_0(J^{P}=0^{-})$, the expected and the observed  $p_0$ value is obtained by integrating the corresponding test-statistic distribution, respectively,  above the $J^{P}=1^{-}$ $q$ distribution median and above the observed value of $q$.
The  exclusion  of  the  $J^P=0^{-}$ hypothesis  in favor of the dark photon $J^P=1^{-}$ hypothesis is  evaluated in terms of the corresponding $\text{CL}_{\textrm{s}}(0^-)$, defined as:
\begin{equation}
\text{CL}_{\textrm{s}}(0^-) = \dfrac{p_0(J^{P}=0^{-})}{1-p_0(J^{P}=1^{-})} \,.
\end{equation}
In the following, we always assume the observed value of the test statistics to be $q=0$.

\subsection{Results}
%
\begin{figure}
\includegraphics[width=0.45\textwidth]{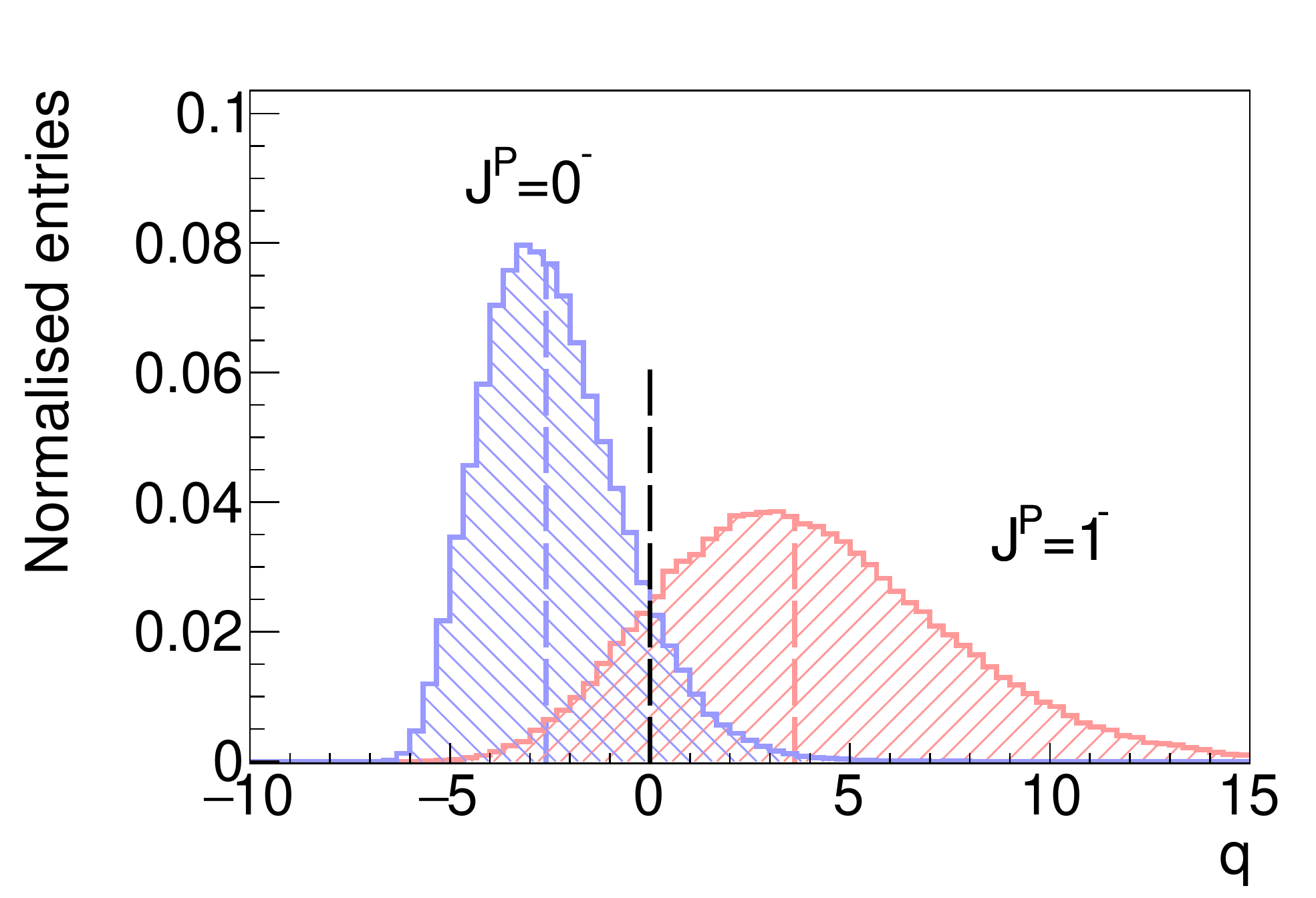}
\caption{\small Expected distributions of the log likelihood ratio test statistics $q$ under the $J^{P}=0^{-}$ and $J^{P}=1^{-}$ spin hypotheses, both for the $J^{P}=0^{-}$  (left) and $J^{P}=1^{-}$ (right) signals. Distributions are obtained using $n_\text{toys}=160000$ and assuming $N=10$ signal events. The expected medians are indicated by vertical dashed lines, and analogously the hypothetical observed value is assumed to occur at $q=0$. 
}
\label{fig:llr}
\end{figure}
The example described in Figure \ref{fig:llr} gives an expected $p_0(J^{P}=0^{-})$ value of $3.9\times 10^{-3}$ and an  observed value of $9.3\times 10^{-2}$. An expected  value $p_0(J^{P}=1^{-})= 2.0\times 10^{-2}$ gives  an expected  exclusion of the $J^{P}=0^{-}$ hypothesis at the 99\% CL, whereas an observed value of $p_0(J^{P}=1^{-})=  1.5\times 10^{-1}$
gives an observed exclusion at 89\% CL. 

Repeating the procedure for all values in the range $N\in\left[2,~25\right]$ we estimate the lower bound for the expected and  observed number of signal events needed to exclude the $p_0(J^{P}=0^{-})$ test hypothesis under the $p_0(J^{P}=1^{-})$ assumption to be, respectively, $N=6$ and $N=17$ at the 95\% CL.
%

 \section{\label{sec:concl}Summary and outlook}
\begin{table*}[t!]
\centering
\renewcommand{\arraystretch}{1.2}
\begin{tabular}{ l c c c c}
 \hline
 &  & & \multicolumn{2}{c}{BR($Z \rightarrow \gamma \dph$)}   \\
 \hline
& $\sqrt{s}$ & $L$ ($\text{ab}^{-1}$) & $M_{T}$ & $E_\gamma$\\ \hline
LHC & $13$ TeV & $0.14$&  $\quad 8 \times 10^{-6}$ & $\quad 5 \times 10^{-5}$\\
\hline 
HL-LHC & $13$ TeV &  $3$ & $\quad 2 \times 10^{-6}$ &  $\quad 1 \times 10^{-5}$\\
\hline
FCC-ee & $91.2$ GeV &  $150$ &  $\quad 2\times 10^{-11}$ & $\quad 3 \times 10^{-11}$\\
\hline
CEPC & $91.2$ GeV & $16$  & $\quad 7 \times 10^{-11}$ & $\quad 8 \times 10^{-11}$\\
\hline 
\end{tabular}
\caption{\small Summary of the upper limits on BR($Z \rightarrow \gamma \dph$) obtained from the simulations performed at the LHC, the HL-LHC and at the FCC-ee and CEPC colliders.}
\label{tabconcl}
\end{table*}

The $Z$-boson decay   into a SM photon and a dark photon would be a most striking signature for  the existence of dark photons. 

The $Z \rightarrow \gamma \dph$ experimental signature   is quite simple and distinctive. In the $Z$-boson CM frame, both the photon and the dark photon  are monochromatic with an energy of  $M_Z/2$. A massless dark photon has 
a neutrino-like signature in a typical experiment, and appears as missing momentum in the $Z \rightarrow \gamma + X$ final state. 

 In this paper we present the estimate of the best exclusion limits on BR($Z\rightarrow \gamma \dph$) for the $Z$ decay into a photon and a dark photon, comparing several present and future collider scenarios: the LHC 
  (at a CM energy of 13 TeV with an integrated luminosity of 140 $\text{fb}^{-1}$), 
 the HL-LHC 
 with 3000 $\text{fb}^{-1}$, and the two future circular leptonic machines FCC-ee and CEPC, at the specific design CM energy of 91.2 GeV ($Z$-factory).
%
%
%
A summary of the 95\% confidence level upper limits on the BR($Z \rightarrow \gamma \dph$)  obtained under the expected conditions of these colliders are collected in Table \ref{tabconcl}. It is then straightforward to compare these results 
with the present LEP bound BR($Z \rightarrow \gamma \dph) < 10 ^{-6}$  at the 95\% CL. 

 The impact of systematic uncertainties at the LHC and the challenge of large QCD backgrounds, intrinsic to hadron colliders,  make all but impossible to  match  the LEP performance. Only a search at the HL-LHC  could  yield a result competing with the LEP limit. 
 
 At the $Z$-factories possibly realized at FCC-ee and/or CEPC, instead,  limits better than  the LEP  one could be obtained,  
 thanks to higher luminosities and efficiencies. The much cleaner environment  could allow a sensitivity to BR($Z \rightarrow \gamma \dph)$ of order $O(10^{-11})$. Such a value comes close to those predicted in dark-sector models  where the effective coupling of the dark photon to the $Z$-boson can be computed.
%
%
\begin{acknowledgments}
{\small
MF is affiliated to the Physics Department of the University of Trieste and the \textit{Scuola Internazionale Superiore di Studi Avanzati} (SISSA), Trieste, Italy. 
MF and EG are affiliated to the Institute for Fundamental Physics of the Universe (IFPU), Trieste, Italy. The support of  these institutions is gratefully acknowledged. }
\end{acknowledgments}

\bibliographystyle{kp.bst}
\bibliography{Zdark}    

\end{document}